\documentclass[aps,prd,twocolumn,nofootinbib,floatfix]{revtex4}
\usepackage{subfigure}
\usepackage{graphicx}
\usepackage{dcolumn}
\usepackage{epsfig}
\usepackage{amsmath}
\usepackage{amsfonts}
\usepackage{amssymb}
\begin{document}

\title{Structural properties of artificial halos in non-standard dark matter simulations}
\author{S. Agarwal and P.-S. Corasaniti}
\affiliation{Laboratoire Univers et Th\'eories (LUTh), UMR 8102, CNRS, Observatoire de Paris, Universit\'e Paris Diderot, 5 Place Jules Janssen, 92190 Meudon, France}

\begin{abstract}
Artificial fragmentation of the matter density field causes the formation of spurious groups of particles in N-body simulations of non-standard Dark Matter (DM) models which are characterized by a small scale cut-off in the linear matter power spectrum. These spurious halos alter the prediction of the mass function in a range of masses where differences among DM models are most relevant to observational tests. Using a suite of high resolution simulations we show that the contamination of artificial groups of particles significantly affect the statistics of halo spin, shape and virial state parameters. We find that spurious halos have systematically larger spin values, are highly elliptical or prolate and significantly deviate from virial equilibrium. These characteristics allow us to detect the presence of spurious halos even in non-standard DM models for which the low-mass end of the mass function remains well behaved. We show that selecting halos near the virial equilibrium provides a simple and effective method to remove the bulk of spurious halos from numerical halo catalogs and consistently recover the halo mass function at low masses.
\end{abstract} 

\maketitle

\section{Introduction}
A generic feature of dark matter models alternative to the standard Cold Dark Matter (CDM) hypothesis is the presence of a small scale cut-off in the linear matter power spectrum. In the Warm Dark Matter (WDM) scenario this arises from the damping of matter density fluctuations on scales smaller than the free-streaming length of WDM particles (see e.g. \cite{Zentner2003,Boyanovsky2008}). Similarly, in models of Ultra-Light Axion (ULA) dark matter, the effective Jean's length associated to the oscillating axion field naturally introduces a damping scale in the matter power spectrum (see e.g. \cite{Hlozek2014}). Scenarios of Late Forming Dark Matter (LFDM) also predict a suppression of power at small scales that depends on the epoch of DM formation \cite{Das2011}. Because of this feature, non-standard DM models and the standard CDM scenario differ in the late-time clustering properties at small scales, where the dynamics of the gravitational collapse is highly non-linear. In this regime quantitative predictions have been possible thanks to the use of large volume high-resolution N-body simulations (see e.g. \cite{Maccio2012,Vogelsberger2012,Lovell2012,Zavala2013,Anderhalden2013,Lovell2014,Abazajian2014,Schneider2014a,Boehm2014,Buckley2014,Agarwal2014}). 

The accuracy of N-body results relies upon the ability to control numerical systematic effects. These can be controlled through convergence analysis tests that evaluate the dependence on the volume and mass resolution of the simulations. However, in the case of models with a sharp cut-off in the initial power spectrum, N-body simulations have shown the presence of unphysical group of particles which result from the artificial fragmentation of the matter density field at small scales \cite{Gotz2002,Gotz2003,WangWhite2007}. These artifacts have been shown to form independently of how initial conditions are generated \cite{WangWhite2007}. This is because artificial fragmentation is a discretization phenomenon which arises from N-body sampling Poisson noise at wave numbers larger than the cut-off in matter power spectrum.

Artificial halos contaminate the low-mass end of the halo mass function in the interval range where difference among the different DM scenarios are most relevant. Hence, their removal is essential to predict the correct abundance of dwarf-galaxy halos as well as the dynamical properties of halo sub-structures that in recent years have become a probe of the nature of DM through observations of the local Universe \cite{BoylanKolchin2011,BoylanKolchin2012,Zavalaetal2009,Papastergis2012,Klypin2014}. 

Numerical studies have shown that increasing the mass resolution of the simulations alleviates artificial fragmentation \cite{WangWhite2007,Schneider2013}. In fact, increasing the number of N-body particles ($N_p$) reduces the intra-particle distance $d$ (i.e. increases the Nyquist frequency of the simulations) as well as the amplitude of the Poisson noise ($\propto 1/N_p$) thus leading to less fragmentation near the cut-off scale. However, it would require the fluid limit to completely remove this effect (for an attempt in this direction see e.g. \cite{Hahn2013,Angulo2013}). Hence, in the presence of artificial halos we ought to rely on empirical methods to infer artifact-free model predictions.

Wang \& White \cite{WangWhite2007} have suggested to apply a mass-cut to numerical halo catalogs and retain only halos with mass larger than $M_{\rm lim}=10.1\,\rho\,d\,k_{\rm peak}^{-2}$, where $\rho$ is the mean cosmic matter density and $k_{\rm peak}$ is the wavenumber of the peak in the dimensionless power spectrum $\Delta^2(k)$ of WDM models and the numerical coefficient is estimated from simulations. 

Schneider, Smith and Reed \cite{Schneider2013} outlined an alternative approach based on the fact that artificial halos contribute to an upturn of the low-mass end of the mass function with a characteristic power law behavior. Using this information, the contribution of artificial halos can be extrapolated at larger masses and subtracted from the measured mass function. 

A more sophisticated approach was adopted by Lovell et al. \cite{Lovell2014}. Their method combines multiple criteria: first, Lagrangian patches in the initial conditions are identified for each halo in the catalogs, those associated to patches whose shape is flatter than a given threshold are removed since genuine proto-halos are spheroidal; second, halos below a given mass-cut are removed; finally, a match between the residual halos in simulations at different resolution is applied and only halos present at both resolutions are retained. Schneider \cite{Schneider2014b} performed a study of the properties of halos in non-standard DM scenarios using a similar approach.

Here, we develop a complementary method that relies on the structural properties of halos as a way to differentiate genuine halos from artificial ones. Using a suite of high resolution N-body simulations of non-standard DM models we show that artificial halos are characterized by extreme values of the spin and shape parameters and significantly depart from virial equilibrium. Hence, discarding unrelaxed halos from numerical halo catalogs can account for most of the effects induced by artificial group of particles and allows to predict the correct halo mass function over a wider range of mass than using simple mass cuts.

The paper is organized as follows: in Section \ref{nbodymodels} we present the models and describe the N-body simulation characteristics. In Section \ref{spuriousresults} we show the results on the halo mass function and the analysis of the structural properties of halos. In Section \ref{haloselection} we propose a halo selection criterion based on the virial state of halos and present the results on the halo mass function and the distribution of halo spins. Finally, we discuss and conclude in Section \ref{conclu}.

\section{N-body Simulations}\label{nbodymodels}
\subsection{Models and Simulation Characteristics}
We run a suite of cosmological simulations using RAMSES \cite{Teyssier2002}, an adaptive mesh refinement code with a tree-based data structure that allows recursive grid refinement on a cell-by-cell basis. Particles are evolved using a particle-mesh (PM) solver, while the Poisson equation is solved using a multigrid method \cite{Guillet2011}.

\begin{figure}
\includegraphics[scale=0.4]{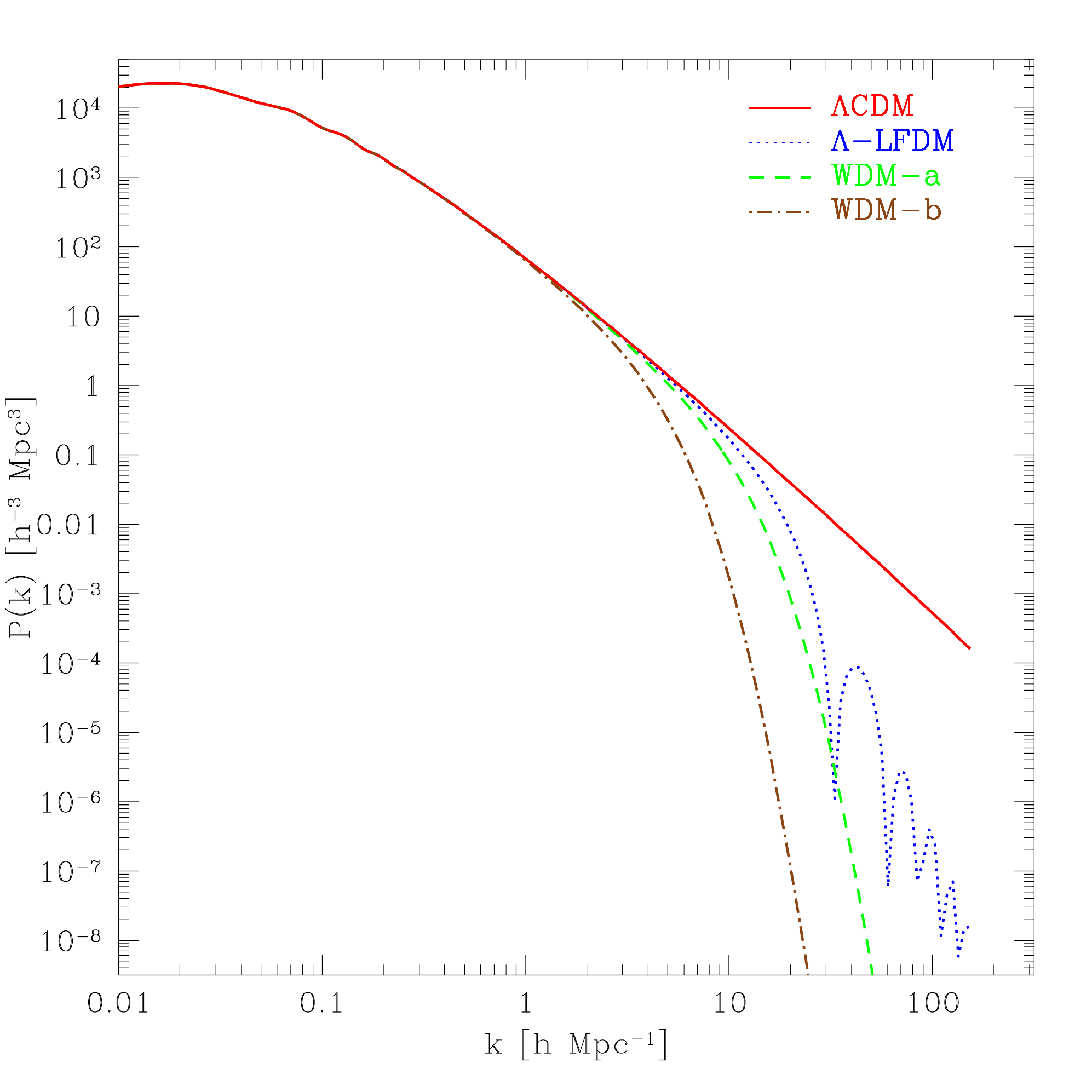}
\caption{Linear matter power spectra at $z=0$ for $\Lambda$CDM (red solid line), LFDM (blue dotted line), WDM-a (green dashed line) and WDM-b (brown dash-dotted line).}\label{fig1}
\end{figure}

We consider as reference model a standard cold dark matter scenario with cosmological constant ($\Lambda$CDM) with zero curvature and the following set of parameters: $\Omega_m=0.3$, h $=0.7$, $\sigma_8=0.8$, $n_s=0.96$ and $\Omega_b=0.046$. For the non-standard DM cosmologies we consider two WDM models with thermal relic particle mass of $m_{\rm wdm}=1.465$ keV (WDM-a) and $0.696$ keV (WDM-b), and a LFDM model with phase transition redshift $z_t=1.5\times 10^6$ (for details on the LFDM model, see \cite{Agarwal2014}). The cosmological parameters for these non-standard DM models are set to our reference $\Lambda$CDM values. We note that out of these three non-standard DM models, only the LFDM satisfies the constraints imposed on the matter power spectrum inferred from the Lyman-$\alpha$ measurements \cite{Viel2013}. WDM models with $m_{\rm wdm}<2$ keV are disfavored at $4\sigma$ confidence level, however we include them in this paper to study artificial fragmentation of the matter density field.
The linear matter power spectra of the non-standard DM models are characterized by a cut-off scale and damping slope\footnote{We compute the initial linear power spectra using CAMB \cite{Lewis2000}. In the case of the WDM models we evaluate the linear transfer function as in \cite{Bode2001}:\begin{equation}
T(k)=\left[1+(\alpha k)^{2\nu}\right]^{-\frac{5}{\nu}},
\end{equation}
with spectrum cut-off scale
\begin{equation}\label{alpha}
\alpha=0.05\left(\frac{\Omega_m}{0.4}\right)^{0.15}\left(\frac{h}{0.65}\right)^{1.3}\left(\frac{\rm 1\,keV}{\rm m_{wdm}}\right)^{1.15}\left(\frac{1.5}{g_{\rm wdm}}\right)^{0.29}
\end{equation}
in units of h$^{-1}$ Mpc. We assume damping slope $\nu=1$ and number of degrees of freedom $g_{\rm wdm}=1.5$. The power spectrum of WDM models is then obtained from that of $\Lambda$CDM as
\begin{equation}
P_{\rm WDM}(k)=T(k)^2P_{\Lambda \rm CDM}(k).
\end{equation}}. These are shown for $z=0$ in Fig.~\ref{fig1}. We can see that the WDM-b model has a larger scale cut-off compared to WDM-a and LFDM, while the damping slope of LFDM is less steep than that of the WDM models. 

We use the linear power spectra to generate Gaussian initial conditions using the Zel'dovich approximation as implemented in MPGRAFIC \cite{Prunet2008}. To facilitate comparison we have used the same phase of the initial conditions for all simulated models. We set the starting redshift of the simulation by imposing the constraint relation $\sigma(\Delta_x^{\rm coarse})=0.02$, such that at the initial redshift all models have the same value of the standard deviation of the initial density field $\sigma$ smoothed on the scale of the coarse grid $\Delta_x^{\rm coarse}$. This guarantees that the initial redshift of the simulations is sufficiently high to suppress spurious effects due to transients \cite{Crocce2006}.

\begin{table} 
\begin{center}
\begin{tabular}{cccc}
\hline\hline
\textrm{Model} & $L$ (h$^{-1}$ Mpc) & $N_\textrm{p}$ & $m_p$ (h$^{-1}$ M$_\odot$) \\
\hline
$\Lambda$CDM & $27.5$ & $512^3$ & $1.29\times 10^7$\\
$\Lambda$CDM & '' & $1024^3$ & $1.61\times 10^6$ \\
LFDM & '' & $512^3$ & $1.29\times 10^7$\\
LFDM & '' & $1024^3$ & $1.61\times 10^6$ \\
WDM-a & '' & $512^3$ & $1.29\times 10^7$\\
WDM-a & '' & $1024^3$ & $1.61\times 10^6$ \\
WDM-b & '' & $512^3$ & $1.29\times 10^7$\\
WDM-b & '' & $1024^3$ & $1.61\times 10^6$\\
WDM-b & 64 & $512^3$ & $1.63\times 10^8$\\
\hline
\end{tabular}
\caption{N-body simulation characteristics. $L$ is the simulation box length, $N_p$ is the number of N-body particles and $m_p$ the mass resolution.\label{tablesim}}  
\end{center}
\end{table}
 
In Table~\ref{tablesim} we list the characteristics of the N-body simulation runs. For all models we run (27.5 h$^{-1}$ Mpc)$^3$ volume simulations with 512$^3$ and $1024^3$ particles corresponding to a mass resolution of $m_p=1.29\times 10^7$ h$^{-1}$M$_\odot$ and $m_p=1.61\times 10^6$ h$^{-1}$M$_\odot$ respectively. For WDM-b we have also run a lower resolution simulation in larger volume: (64 h$^{-1}$ Mpc)$^3$ volume with $512^3$ particles corresponding to a mass resolution $m_p=1.63\times 10^8$ h$^{-1}$M$_\odot$. This simulation suite allows us to perform a detailed convergence study on artificial halos.

\begin{figure*}[ht]
\begin{centering}
\subfigure{\includegraphics[scale=0.4]{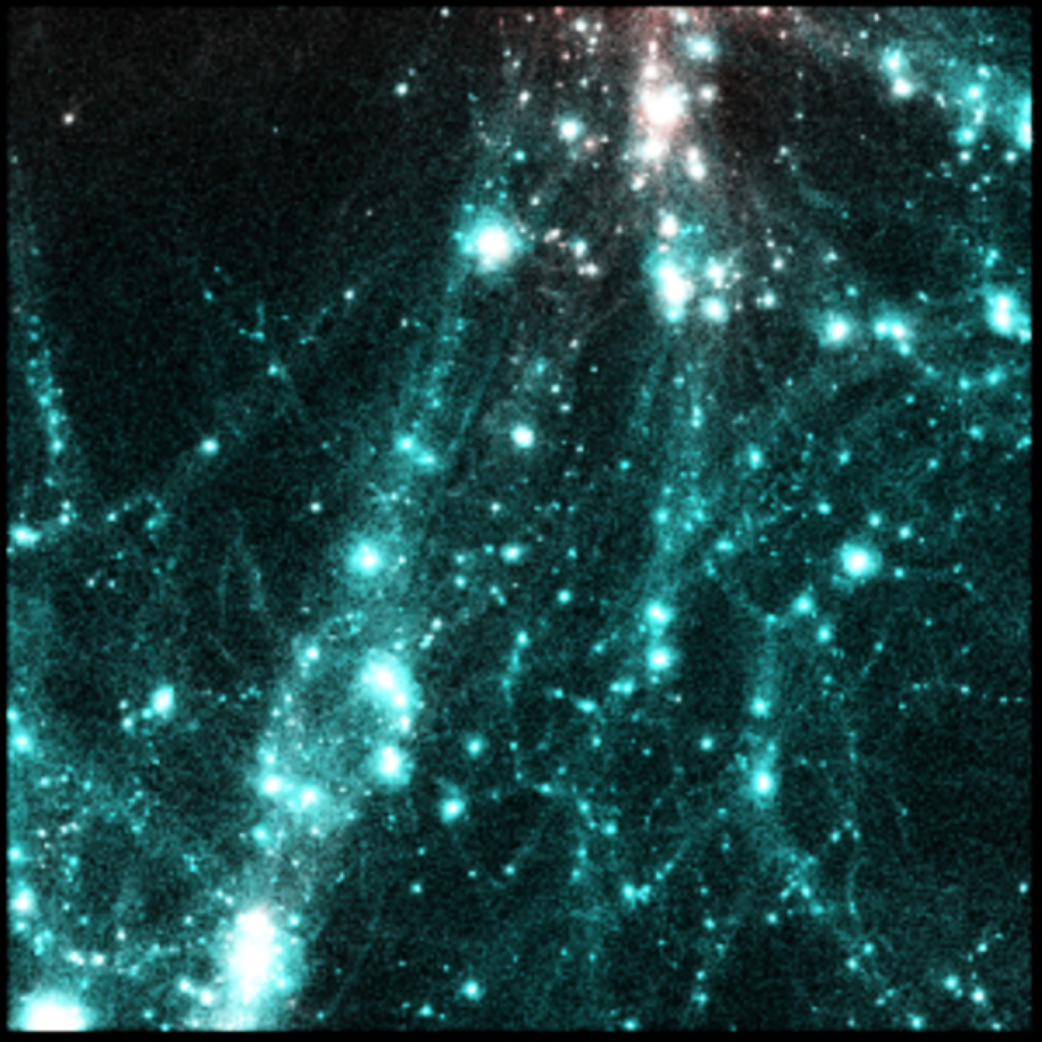}}\subfigure{\includegraphics[scale=0.4]{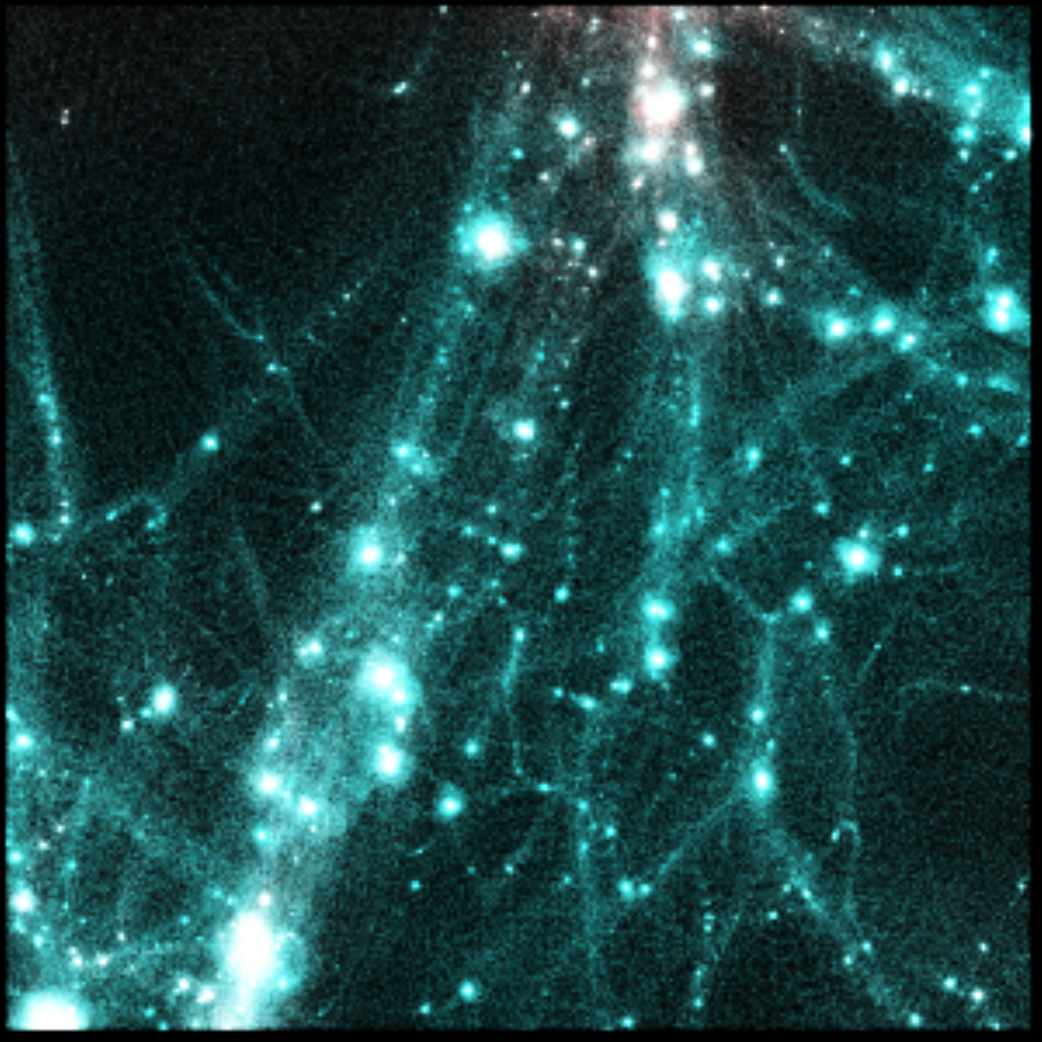}\includegraphics[scale=0.4]{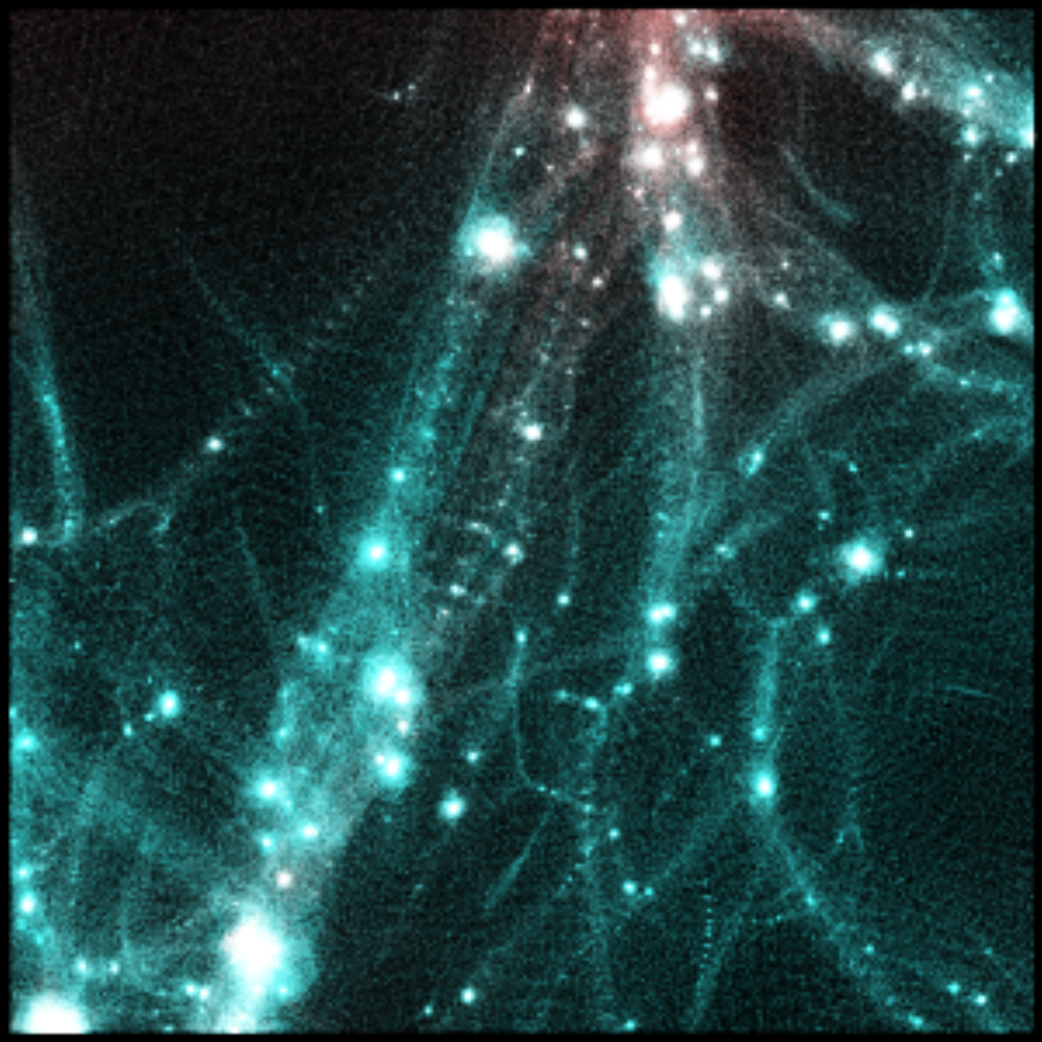}}\subfigure{\includegraphics[scale=0.4]{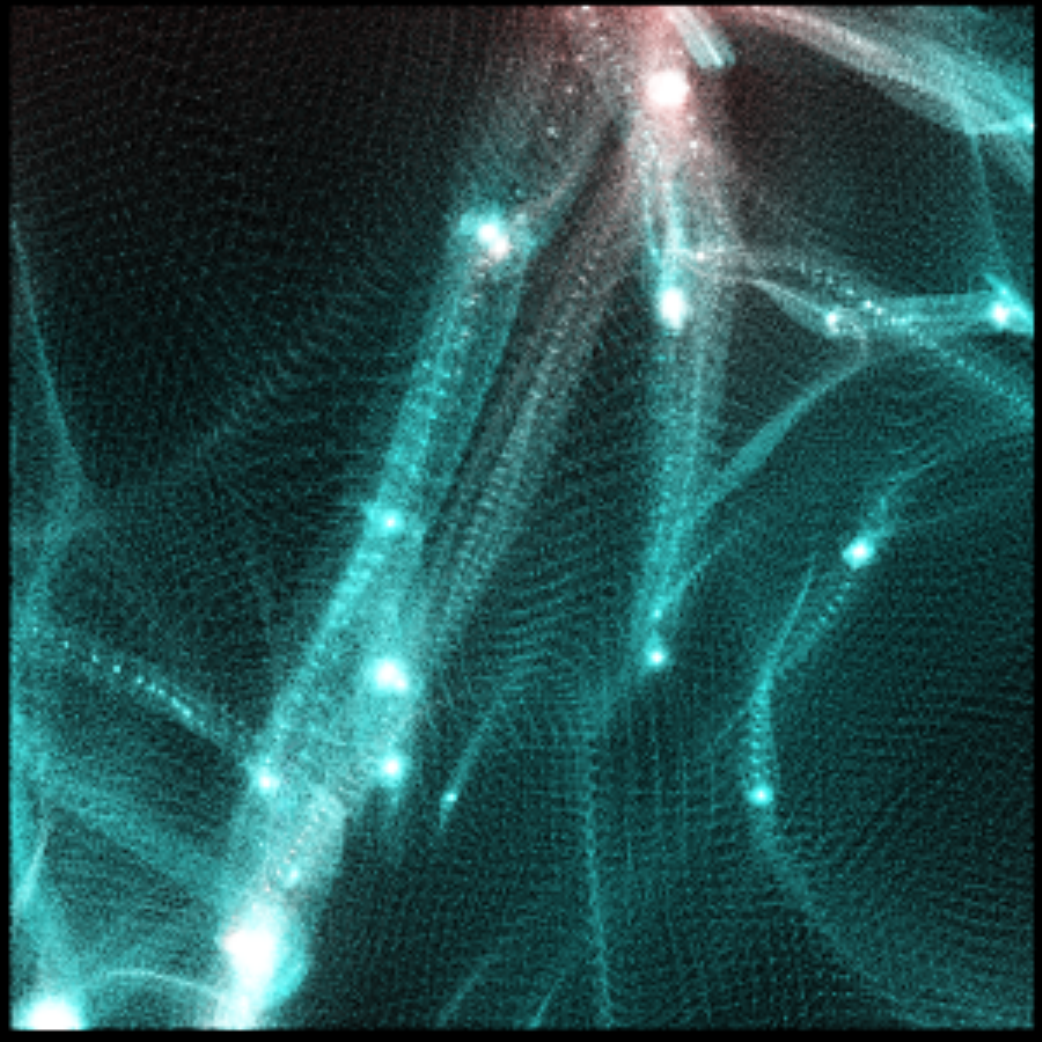}}
\caption{2D projection of the dark matter density field at $z=0$ in a box of $\sim 7$ h$^{-1}$Mpc side from the simulations with (27.5 h$^{-1}$ Mpc)$^3$ volume and 512$^3$ particles for $\Lambda$CDM, LFDM, WDM-a and WDM-b (from left to right).}\label{fig2}
\end{centering}
\end{figure*}

\subsection{Halo Finder and Halo Properties}
We detect halos using the code pFOF \cite{Roy2014} based on a parallelized version of the friend-of-friend algorithm \cite{Davis1985}. This detects halos as group of particles characterized by an intra-particle distance smaller than a given linking length parameter $b$, which we set to $0.2$. During the first iteration, the algorithm groups all particles in the simulation box which are within distance $b$ of an initial particle. This search is then carried out for the rest of the particles of the group until no new neighbors are found. This final set of particles is then tagged as a halo and removed from the list. The algorithm then iterates over the next untagged particles repeating the above procedure to detect a new halo. 

For each halo in the simulations we estimate the spin parameter defined as \cite{Bullock2001}:
\begin{equation}\label{lambdaprime}
\lambda'=\frac{J}{\sqrt{2}MVR},
\end{equation}
where $M$ is the mass of the halo, $V=\sqrt{GM/R}$ is the orbital velocity at the virial radius $R$ ($G$ being Newton's constant) and $J$ is the amplitude of the halo angular momentum:
\begin{equation}
\vec{J}=m_p \sum_{i=1}^{N_h} (\vec{r}_i-\vec{r}_c)\times(\vec{v}_i-\vec{v}_c),
\end{equation}
where the sum runs over all $N_h$-halo particles with position $\vec{r}_i$ and velocity $\vec{v}_i$ ($\vec{r}_c$ and $\vec{v}_c$ being the position and velocity of the halo center of mass). The definition of the spin parameter given by Eq.~(\ref{lambdaprime}) normalizes the halo angular momentum to its maximal value.

We also quantify the shape of halos by computing the symmetric mass distribution tensor defined as (see e.g. \cite{Thomas1998})
\begin{equation}\label{shapetensor}
M_{\alpha\beta}=\frac{m_p}{M}\sum_{i=1}^{N_h} (r_{\alpha,i}-r_{\alpha,c})(r_{\beta,i}-r_{\beta,c}),
\end{equation}
where $\alpha,\beta=1,2,3$ denote the three components of the position vectors. The eigenvalues of $M_{\alpha\beta}$, $a^2\ge b^2 \ge c^2$, define a triaxial ellipsoid with axis lengths $a \ge b \ge c$. These can be combined to define halo shape parameters. Here, we focus on the sphericity $s=c/a$, ellipticity $e=\frac{1}{2}(a-c)/(a+b+c)$ and prolateness $p=\frac{1}{2}(a-2b+c)/(a+b+c)$. Genuine halos, being close to spheroidal, have large values of sphericity and low ellipticity and prolateness. On the other hand, spurious halos are expected to be abnormally elliptical (large value of $e$) with disk-like (oblate, $p<0$) or needle-like (prolate, $p>0$) shapes.

We estimate the dynamical state of halos by computing the \textit{virial state} parameter $\eta = 2K/|E| \in [0,\infty]$, where $K$ is the total kinetic energy and $E$ is the gravitational potential energy. In the case of virialized halos $\eta\approx 1$. It is worth reminding that $\eta$ only provides an approximate estimation of a halo's virial state. In fact, while there is no ambiguity in the determination of the kinetic energy of the halo (i.e. the sum of the kinetic energy of each halo particle), the potential energy $E$ is a non-local quantity, since it also depends on particles that do not belong to the halo, but are in the surrounding density field. This may introduce a systematic source of uncertainty in the evaluation of the virial state. Despite such a limitation, $\eta$ remains a useful proxy especially to identify halos with large deviation from the virial condition (i.e. $\eta\gg 1$).

\section{Analysis of Spurious Halos}\label{spuriousresults}
A visual inspection of the simulations shows that there are progressively less structures in LFDM, WDM-a and WDM-b compared to the $\Lambda$CDM. This can be seen in Fig.~\ref{fig2} where we plot a 2D projection of the dark matter density field at $z=0$ in a box of $\sim 7$ h$^{-1}$Mpc side from the (27.5 h$^{-1}$ Mpc)$^3$ volume simulations with 512$^3$ particles. This trend is consistent with expectations due to the fact that the damping of the linear matter power spectrum of the non-standard DM models considered here occurs at an increasingly larger scale (small wavenumbers) as we go from LFDM to WDM-b (see Fig.~\ref{fig1}). For these models we can also notice an increasing granularity of the density field along filamentary structures with the appearance of regularly spaced group of particles that are especially distinguishable in the WDM-b case. These are spurious numerical halos. 

\subsection{Halo Mass Function}
The presence of a cut-off in the linear matter power spectrum introduces a characteristic mass scale in the halo mass function. As shown in \cite{Schneider2012} for WDM models, this scale can be parameterized in terms of the `half-mode' mass M$_{\rm hm}$, namely the mass associated to the length scale where the linear transfer function drops to half of its value. In the case of the WDM models considered here we have M$_{\rm hm}=7.41\times 10^9$ h$^{-1}$ M$_\odot$ for WDM-a and M$_{\rm hm}=9.65\times 10^{10}$ h$^{-1}$ M$_\odot$ for WDM-b, while for the LFDM model we estimate M$_{\rm hm}=3\times 10^9$ h$^{-1}$ M$_\odot$.

\begin{figure}[th]
\includegraphics[scale=0.4]{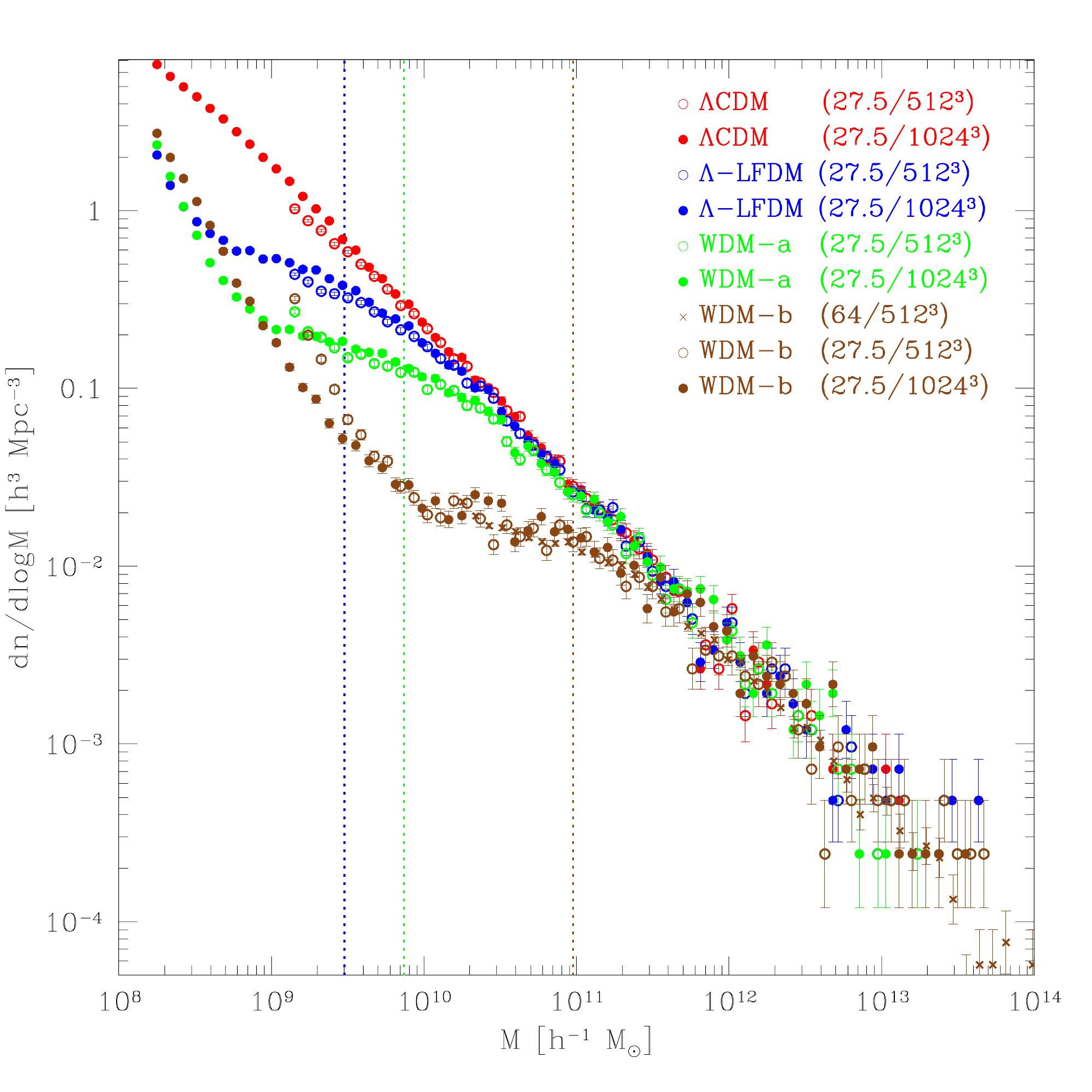}
\caption{Halo mass function at $z=0$ for $\Lambda$CDM (red circles), LFDM (blue circles), WDM-a (green circles) and WDM-b (brown circles/stars) from the simulations listed in Table~\ref{tablesim}. Errorbars are given by Poisson errors. Vertical dotted lines indicate the value of the half-mode mass for LFDM (blue), WDM-a (green) and WDM-b (brown) models.}\label{fig3}
\end{figure}

In Fig.~\ref{fig3} we plot the halo mass function at $z=0$ for halos with at least 100 particles from the simulation suite. Let us first consider the results of the (27.5 h$^{-1}$ Mpc)$^3$ volume runs with 512$^3$ particles. We can see that the number density of halos in the non-standard DM models is suppressed with respect to the $\Lambda$CDM case at around the half-mode mass as expected. The larger the half-mode mass the larger the suppression of the mass function relative to $\Lambda$CDM. However, in the case of the WDM models we can see that the low mass end of the mass function rises instead of falling. The effect is more dramatic for WDM-b model which has the largest half-mode mass. As shown in \cite{WangWhite2007} this upturn is an artifact due to the contribution of artificial halos. Above the half-mode mass all models converge to the $\Lambda$CDM mass function within finite volume errors. Because of the absence of an upturn in the LFDM case one might think that the model is exempt from spurious halo contamination. However, the mass function from the higher resolution run shown in Fig.~\ref{fig3} reveals this not to be the case since an upturn is present at much smaller masses.

Let us compare the mass function from the (27.5 h$^{-1}$ Mpc)$^3$ volume simulations with $512^3$ and $1024^3$ particles. From Fig.~\ref{fig3} we see that the slope of the upturn at the low-mass end depends on the mass resolution of the simulations, in agreement with the findings of \cite{Schneider2013}. This dependence is a clear indication of the fact that the majority of halos in this mass range are numerical artifacts. We find that the mass functions of the lower resolution runs converge to those of higher resolution at $\approx 5\%$ for M$\gtrsim 4\times 10^9$ h$^{-1}$ M$_\odot$ -- roughly corresponding to halos with more than $300$ particles. For the WDM-b simulation of (64 h$^{-1}$ Mpc)$^3$ volume and $512^3$ particles we find a similar level of convergence  for M$\gtrsim 5\times 10^{10}$ h$^{-1}$ M$_\odot$, again roughly corresponding to a minimum of $300$ particles per halo. Thus, from now on we only consider halos with at least $300$ particles. For simulations of (27.5 h$^{-1}$ Mpc)$^3$ volume and $512^3$ particles, a $300$-particle halo has a mass of $4\times 10^9$ h$^{-1}$ M$_\odot$ and well below the half-mode mass M$_{\rm hm}$ of the WDM models (green and blue vertical dotted lines in Fig.~\ref{fig3}). The WDM mass functions, instead of falling, continue to rise even below their respective M$_{\rm hm}$ values. As such, even a conservative $300$-particle cut does not fully solve the problem of spurious halos. Moreover, from the analysis of the LFDM mass function with $512^3$ particles we can deduce that for a given mass resolution simulation the absence of a well defined upturn at the low-mass end of the mass function does not imply that a model characterized by a suppression of power at small scales is exempt from artificial halo contamination.

The point that we want to stress is that artifacts caused by spurious halos cannot be addressed by solely assuming conservative mass cuts in the halo catalogs. Moreover, extrapolating information from the upturn of the mass function might not be possible since depending on the cosmological model and the simulation characteristics an upturn may be absent near the half-mode mass. A case in point is the LFDM mass function (see Fig.~\ref{fig3}) from the (27.5 h$^{-1}$ Mpc)$^3$ volume simulations with $512^3$ particles, where at the lowest mass end there is no well defined upturn. As we will show hereafter, removing the contribution of artificial halos requires their characterization in terms of physical properties that may distinguish them from genuine halos.

\begin{figure}[ht]
\includegraphics[scale=0.4]{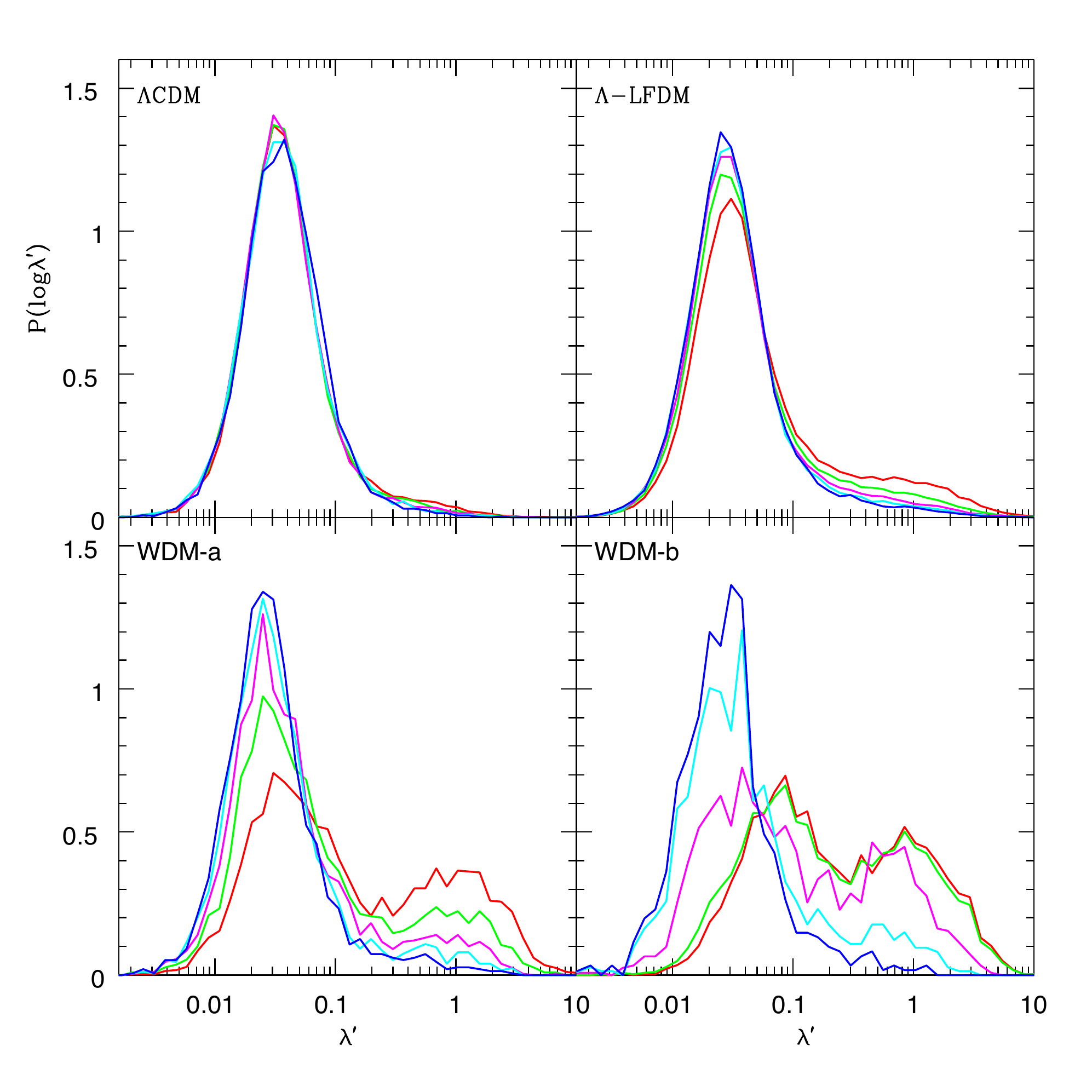}
\caption{Probability density function of the spin parameter at $z=0$ for $\Lambda$CDM (top left panel), LFDM (top right panel), WDM-a (bottom left panel) and WDM-b (bottom right panel). The different lines correspond to $5$ equally spaced logarithmic mass bins in the range $4<\textrm{M}[10^9\,\textrm{h}^{-1}\,\textrm{M}_\odot]<8$ for the LFDM, $\Lambda$CDM and WDM-a models and $4<\textrm{M}[10^9\,\textrm{h}^{-1}\,\textrm{M}_\odot]<100$ for the WDM-b model. The red (blue) solid line corresponds to the lowest (highest) mass bin.}\label{fig4}
\end{figure}

\begin{figure}
\begin{tabular}{c}
\includegraphics[scale=0.4]{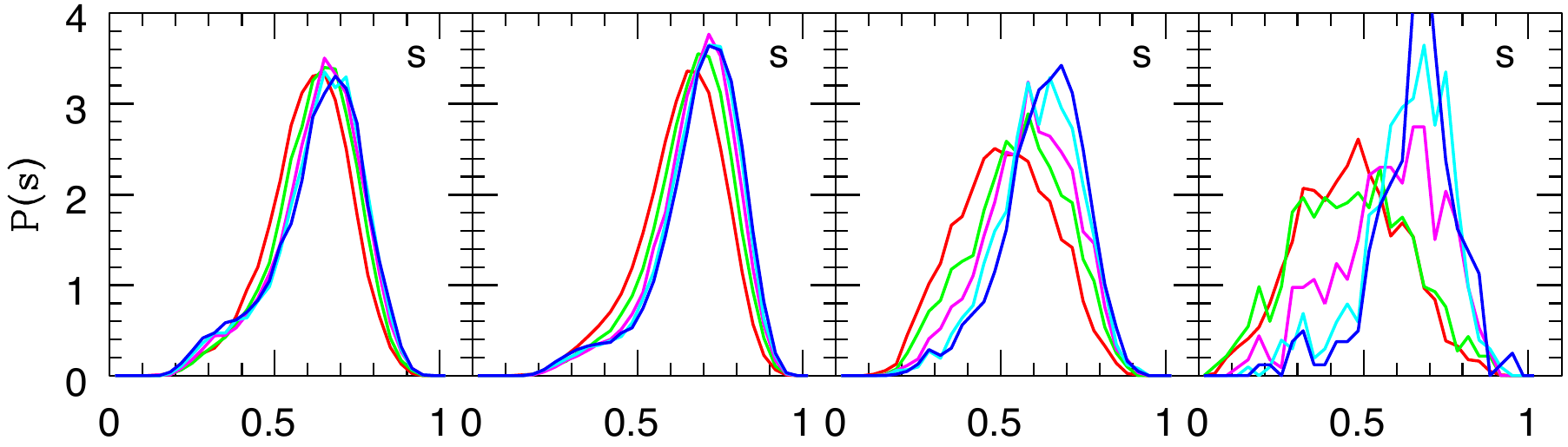}\\
\includegraphics[scale=0.4]{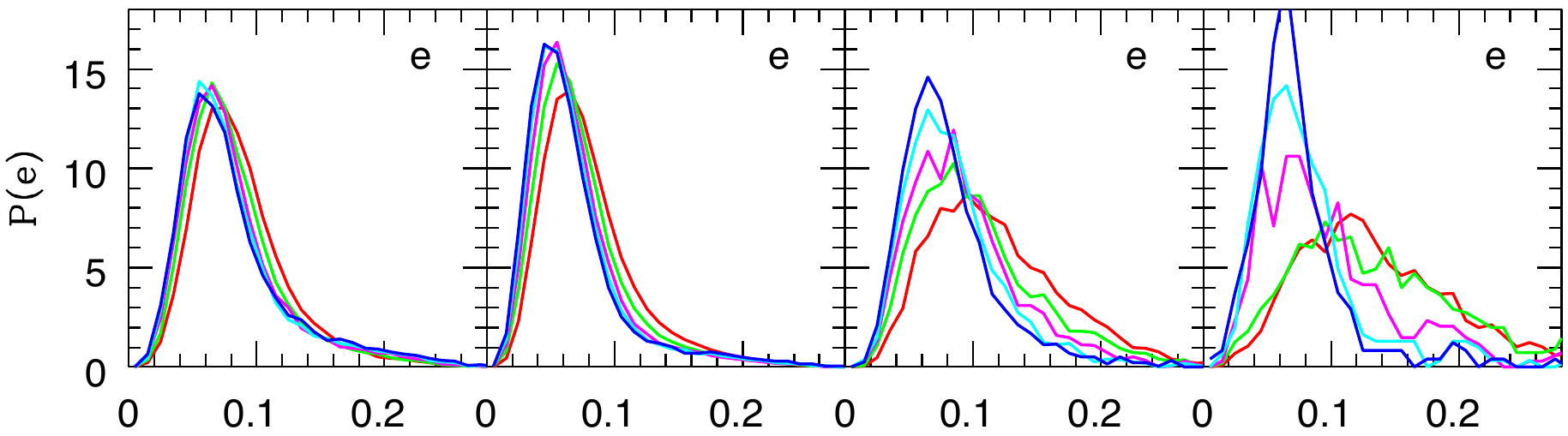}\\
\includegraphics[scale=0.4]{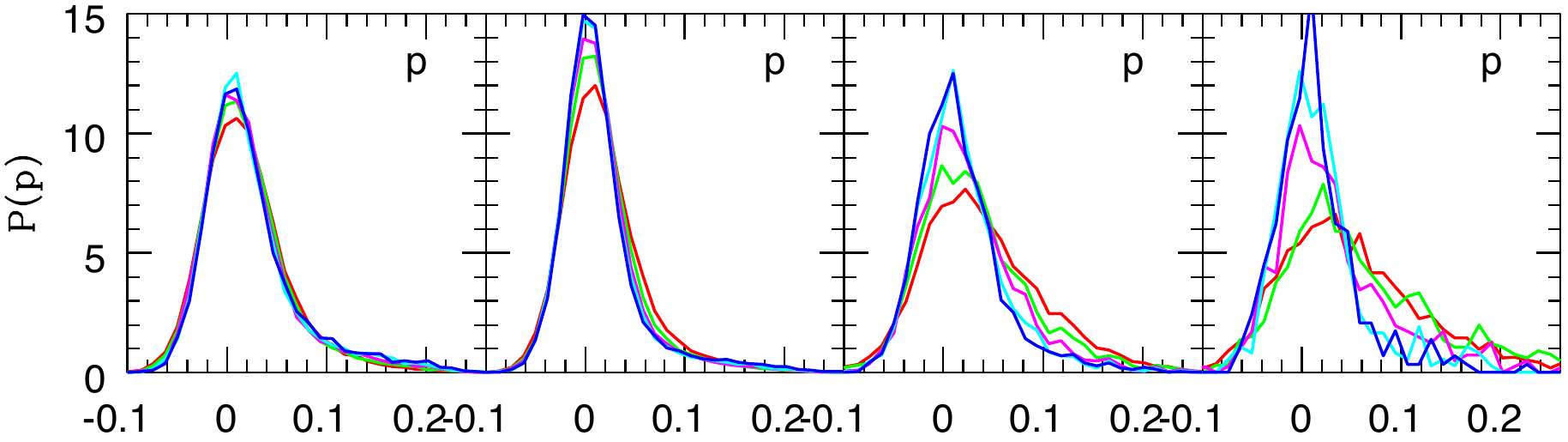}
\end{tabular}
\caption{Probability density function of halo sphericity, ellipticity and prolateness (panels top to bottom) at $z=0$ for $\Lambda$CDM, LFDM, WDM-a and WDM-b (panels left to right) respectively. We consider the same mass bins as in Fig.~\ref{fig4}.}\label{fig5}
\end{figure}

\subsection{Halo Spin and Shape}
In Fig.~\ref{fig4} we plot the normalized probability density function of the halo spin parameter at $z=0$ from the boxlength $L=27.5\,{\rm h^{-1} Mpc}$ simulations with $512^3$ particles. The different lines correspond to equally spaced logarithmic mass bins in the range $4<\textrm{M}[10^9\,\textrm{h}^{-1}\,\textrm{M}_\odot]<8$ for $\Lambda$CDM, LFDM and WDM-a, and $4<\textrm{M}[10^9\,\textrm{h}^{-1}\,\textrm{M}_\odot]<100$ for WDM-b\footnote{The upper limit of the mass range roughly corresponds to the largest half-mode mass of the pair of DM models considered.}

Previous studies have shown that the distribution of halo spins in $\Lambda$CDM model depends only weakly on the halo mass and is approximately described by a log-normal distribution \cite{Bett2007,Knebe2008}. As we can see from the top left panel of Fig.~\ref{fig4} our results are consistent with these findings. Notice that the tail of the $\Lambda$CDM distribution appears to be heavier in the lowest mass bin (red solid line). As shown in \cite{Bett2007} this is mostly due to unrelaxed halos. We will confirm this conclusion in the analysis presented in Section~\ref{virialcond}. 

In the top right panel of Fig.~\ref{fig4} we plot the distribution of spins for the LFDM model. Compared to the $\Lambda$CDM case, we can see a stronger dependence on halo mass. The distribution appears to be well described by a log-normal only for the highest mass bin (blue solid line), while becoming increasingly heavy tailed at lower masses.

\begin{figure*}[ht]
\begin{centering}
\subfigure{\includegraphics[scale=0.35]{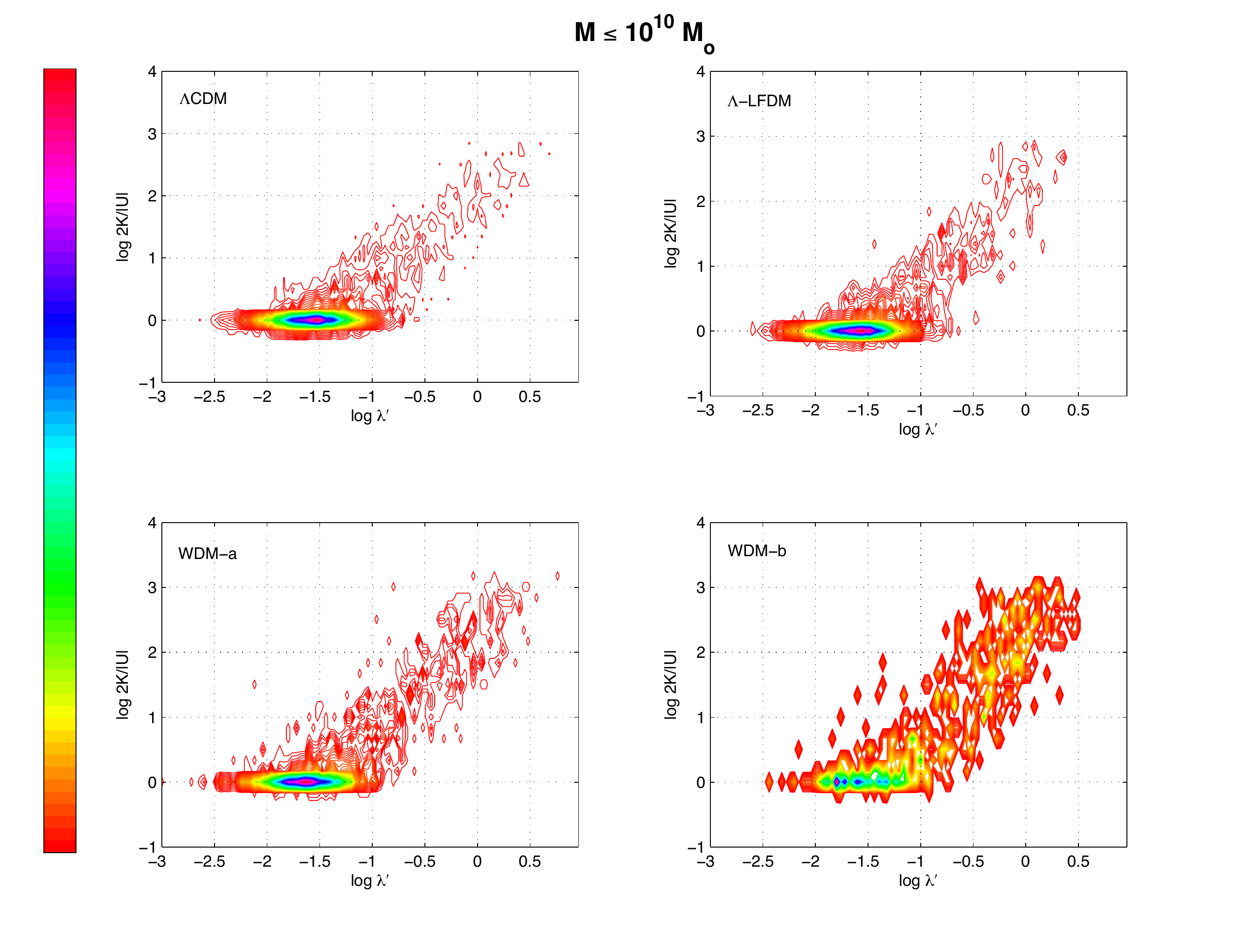}}\subfigure{\includegraphics[scale=0.35]{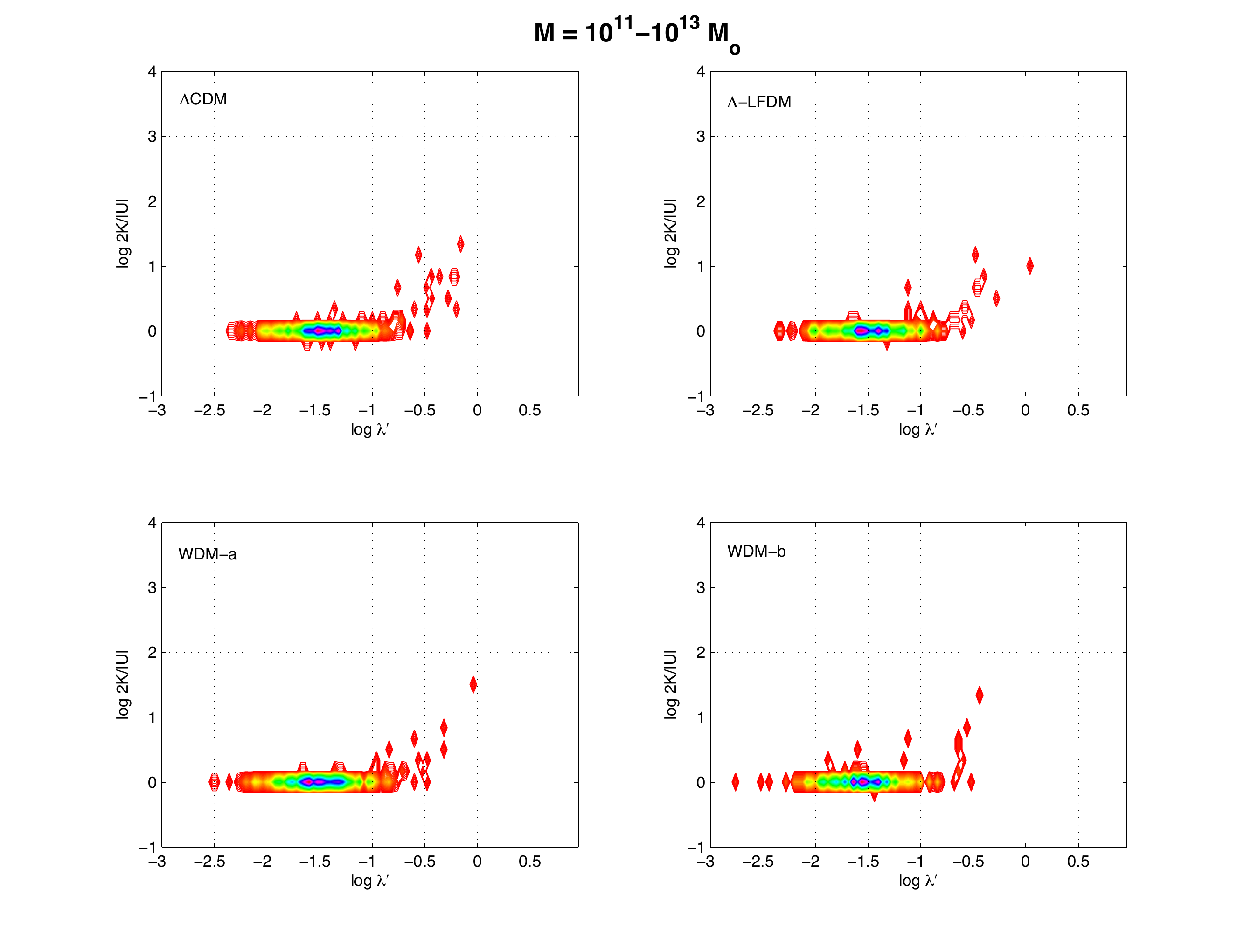}}
\caption{Normalized probability density function in the $\lambda'-\eta$ plane at $z=0$ for halos with mass M$<10^{10}$ h$^{-1}$ M$_\odot$ (left panels) and $10^{11}<\textrm{M}[\textrm{h}^{-1}\textrm{M}_\odot]<10^{13}$ (right panels) in the case of $\Lambda$CDM, LFDM, WDM-a and WDM-b models.}\label{fig6}
\end{centering}
\end{figure*}

Despite a larger statistical noise, such a trend can also be seen in the WDM models shown in the lower panels of Fig.~\ref{fig4}. We may notice large departures from log-normality. As in the LFDM case the distribution is increasingly heavy tailed, in particular the lower mass bins are characterized by a bimodal distribution of the spin parameter. Since spurious halos dominate the halo abundance in this mass range (see Fig.~\ref{fig3}), we can deduce that the bimodality is a numerical artifact similar to the upturn in the halo mass function. Furthermore, as the lowest mass bin in the LFDM case in Fig.~\ref{fig4} shows a heavy tail compared to a log-normal distribution, this indicates that spurious halos also affect the LFDM halo mass function (see Fig.~\ref{fig3}) as confirmed by the presence of an upturn in the higher resolution run. This implies that other non-standard DM scenarios characterized by a suppression of power at small scales similar to LFDM, such as a class of Self-Interacting Dark Matter (SIDM) models discussed in \cite{Agarwal2014}, may be contaminated by spurious halos as well.

In Fig.~\ref{fig5} we plot the normalized probability density function of halo shape parameters at $z=0$ as a function of halo mass. We can see a mass trend similar to that of the spin parameter shown in Fig.~\ref{fig4}. In the $\Lambda$CDM case the majority of halos are slightly elliptical, moreover the distribution of the shape parameters is nearly independent of the halo mass. This is not the case for the low mass bins of the non-standard DM models. In particular, in the WDM case we see that only the higher mass bins have a distribution that is consistent with that of the $\Lambda$CDM. At lower masses, halos have systematically higher values of ellipticity and prolateness. This suggests that besides high spin values, artificial halos are highly elliptical and prolate compared to genuine ones. Though not directly comparable, these results are consistent with the analysis by Lovell et al. \cite{Lovell2014} which found that in WDM simulations spurious subhalos are associated to proto-halos characterized by small values of the sphericity, while genuine subhalos at higher masses are more spheroidal and closer to that of $\Lambda$CDM prediction. This trend is similar to that of the sphericity distribution shown in the top panel of  Fig.~\ref{fig5}. Similar results were found in \cite{Schneider2014b} which studied the distribution of shape parameters in non-standard DM scenarios with power suppression at small scales.

\subsection{Virial State vs Spin Statistics}\label{virialcond}
Further insight on the structural properties of artificial halos can be gained by considering the probability distribution of the virial state parameter $\eta$. Since spurious halos are the result of artificial fragmentation, it is reasonable to expect that such group of particles are not virialized. Therefore, in the case of artificial halos the parameter $\eta$ deviates from unity and correlates with large values of halo spin. 

In Fig.~\ref{fig6} we show the density plot of the normalized probability density function in the $\lambda'-\eta$ plane for the simulations with $512^3$ particles and boxlength $L=27.5\,{\rm h^{-1} Mpc}$ $z=0$. The set of four panels on the left shows the distribution of halos with mass M$<10^{10}$ h$^{-1}$M$_\odot$, while the set on the right shows halos with mass in the range $10^{11}<\textrm{M}[\,\textrm{h}^{-1}\,\textrm{M}_\odot]<10^{13}$. In the former case we can clearly see a strong correlation between large deviations from virial equilibrium (i.e. $\eta=1$) and high spin values. This manifests through a heavy tail that is absent in the higher mass bin case. From the density plot we see that the fraction of halos populating this extended tail is lowest for $\Lambda$CDM, while it increases for the non-CDM models in proportion to their half-mode mass. Moreover, the tail systematically extends towards larger values of $\lambda'$ and $\eta$. We also notice that in the case of the WDM models the tail of the distribution becomes bimodal. Let us remark that while the $\Lambda$CDM halos with M$<10^{10}$ h$^{-1}$M$_\odot$ shown Fig.~\ref{fig6} have a distribution of spin and shape parameters which is nearly independent of their mass (see Figs.~\ref{fig4}-\ref{fig5}), this is not the case for the non-standard DM models. This suggests that while the extended tail of the $\Lambda$CDM probability density function in the $\lambda'-\eta$ plane is composed of genuine unrelaxed halos, that of the non-CDM models is dominated by spurious halos.

\section{Spurious Halos Selection}\label{haloselection}
In the previous Section we have shown that the structural properties of spurious halos in non-standard DM simulations differ from that of genuine halos\footnote{Here, we have not considered the density profile of halos, since spurious groups of particles are irregularly shaped and unrelaxed. As shown in \cite{Balmes2014} the density profile of such perturbed systems is badly fit by the universal Navarro-Frenk-White profile, consequently the concentration parameter is completely uninformative about the internal structure of the halos.}. In particular, we have found a strong correlation between deviations from the virial condition and large values of halo spin at low masses. Spurious halos with such features contribute to the heavy tale of the spin distribution in the LFDM model and the bimodality in WDM models. These results suggest a simple way of removing artificial halos from N-body halo catalogs since a cut based on the virial state of halos, as parametrized by $\eta$, can filter out spurious halos in the tail of the halo distribution. In the CDM case this approach leaves only halos that are approximately well relaxed (see e.g. \cite{Bett2007}), while it effectively removes the bulk of spurious halos in non-CDM models. As we will show here, this halo selection allows us to recover the halo mass function consistent with expectations at low masses and gives convergent results when applied to N-body runs with different mass resolution.

\begin{figure}[ht]
\includegraphics[scale=0.4]{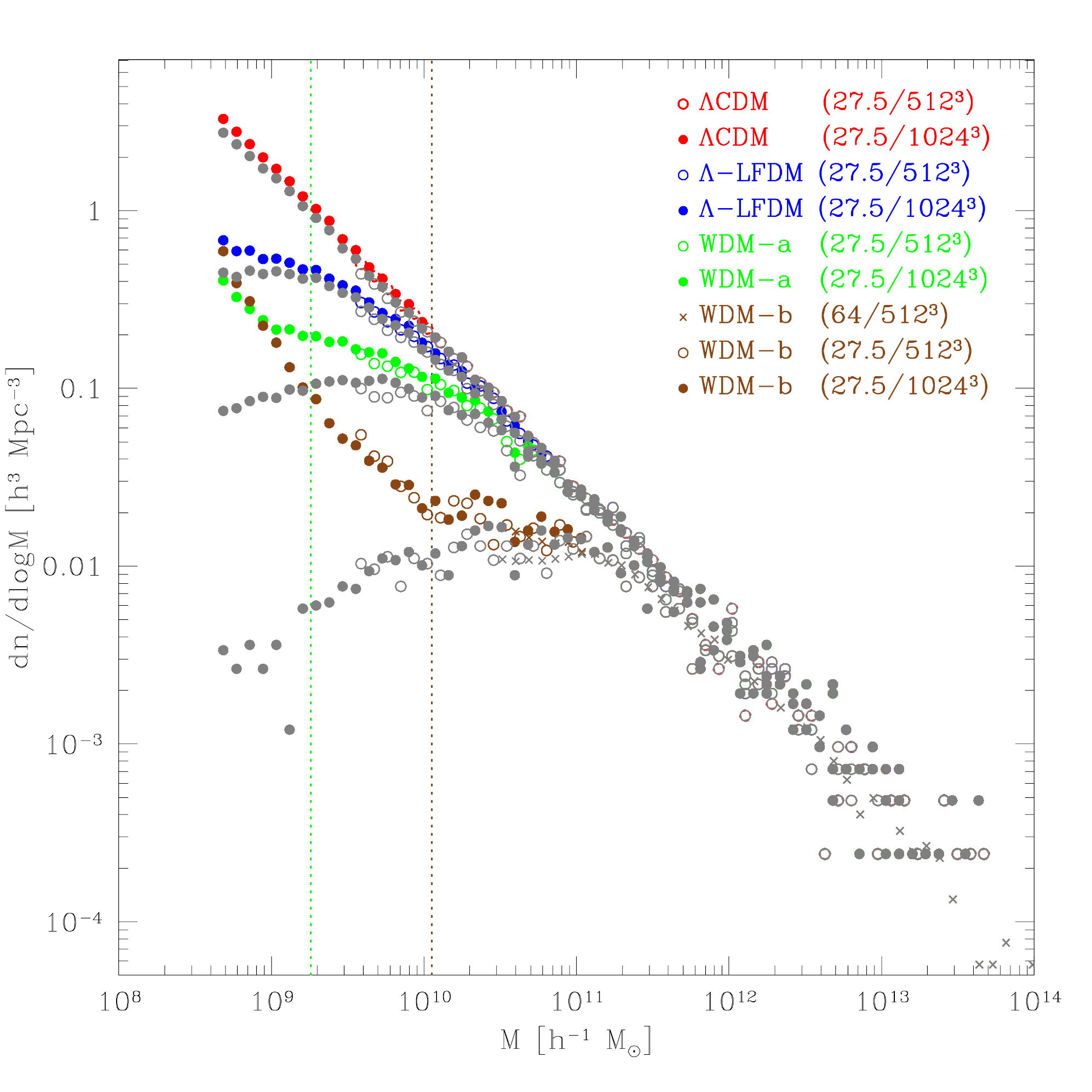}
\caption{Halo mass function at $z=0$ for $\Lambda$CDM (red circles), LFDM (blue circles), WDM-a (green circles) and WDM-b (brown circles/stars) from the simulations listed in Table~\ref{tablesim} before and after removing unrelaxed halos (grey points). All halos have at least $300$ particles. Vertical lines correspond to the mass cut $M_{\rm lim}$ from \cite{WangWhite2007} for WDM-a (green) and WDM-b (brown) models in the case of the simulations with boxlength $L=27.5\,{\rm h^{-1} Mpc}$ and $512^3$ particles.}\label{fig7}
\end{figure}

In Fig.~\ref{fig7} we plot the halo mass function at $z=0$ for halos with at least $300$ particles for the simulations listed in Table~\ref{tablesim}. The grey points show the mass functions for halos with the virial state parameter in the range $0\leq \eta \leq1.5$. In particular, let us focus on the mass function of LFDM (blue points), WDM-a (green points) and WDM-b (brown points) models before and after the virial state parameter cut from the simulations of boxlength $L=27.5\,{\rm h^{-1} Mpc}$ with $512^3$ (empty circles) and $1024^3$ (solid circles) particles. Notice that discarding unrelaxed halos ($\eta>1.5$) has removed the upturn at low masses. Now, the mass functions drop as expected below the half-mode mass of each model. Moreover, we can see that for the LFDM and WDM-a models, in the mass range where the halo catalogs from the low and high resolution runs overlap, the respective mass functions converge at $\sim 5\%$ level, while for the WDM-b case a scatter of $\sim 20\%$ is found only in certain mass bins due to Poisson noise. Hence, this approach allows us to recover a numerically convergent mass function over a wider range of low masses compared to the aggressive $M_{\rm lim}$ cut \cite{WangWhite2007} that discards all halos -- genuine or otherwise -- with mass $M<M_{\rm lim}$ while still leaving the halo abundance uncorrected between $M_{\rm lim}$ and $M_{\rm hm}$. This suggests that the halo selection criterion proposed here can efficiently remove spurious halos (i) independently of the resolution of the simulations, and (ii) independently of the specificities of the cosmological model.

\begin{figure}[ht]
\includegraphics[scale=0.4]{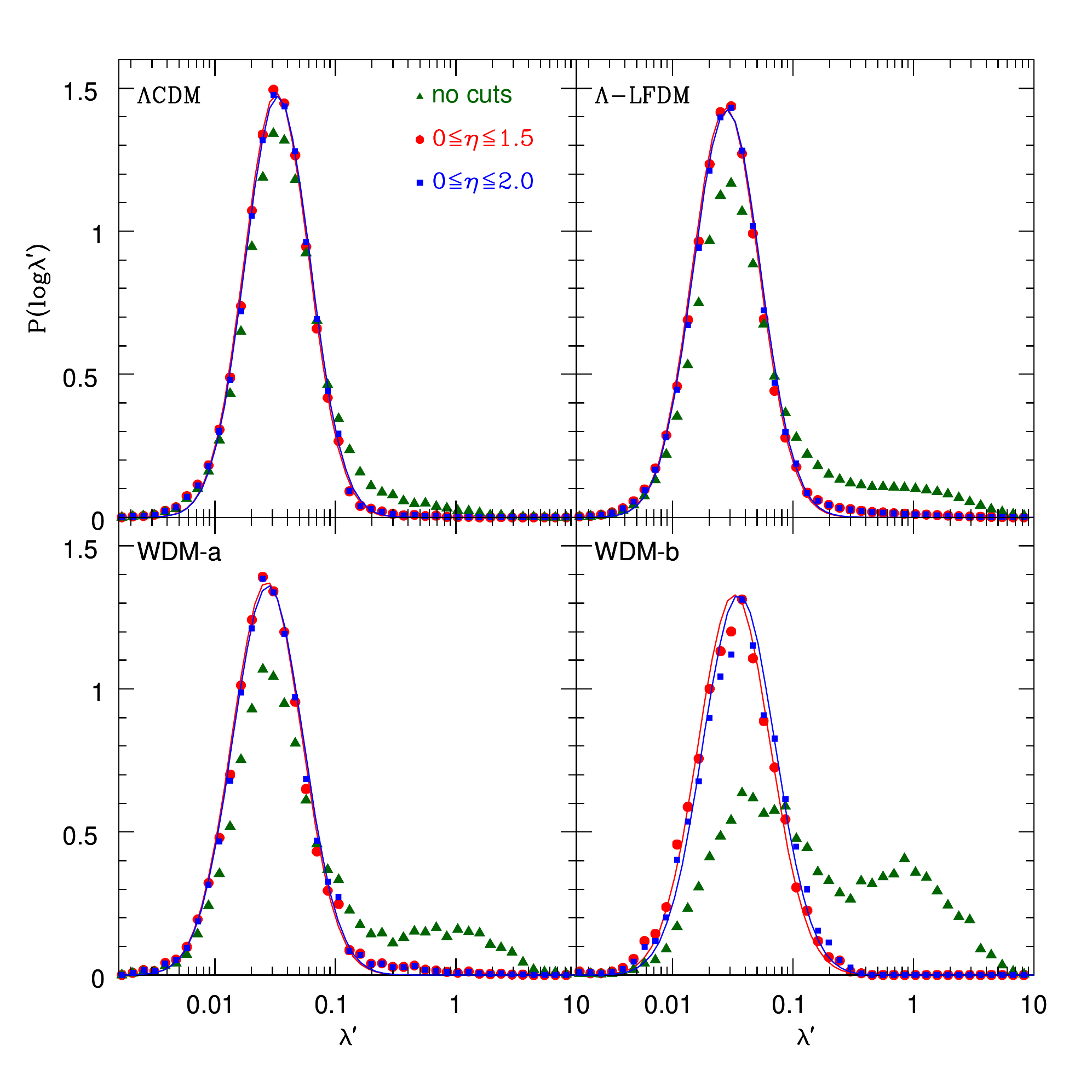}
\caption{Probability density function of the spin parameter at $z=0$ for $\Lambda$CDM (top left), LFDM (top right), WDM-a (bottom left) and WDM-b (bottom right) for all halos in the mass bins shown in Fig.~\ref{fig4}. Green triangles correspond to the distributions inferred before removing unrelaxed halos, while the red circles and blue squares correspond to relaxed halos with $0\leq \eta \leq1.5$ and $0\leq \eta \leq2$ respectively. The red and blue solid curves are the corresponding best-fitting log-normal functions.}\label{fig8}
\end{figure}

In Fig.~\ref{fig8} we plot the normalized probability density function of the spin parameter for all halos shown in Fig.~\ref{fig4} before and after removing unvirialized halos for two different cuts, $\eta>1.5$ and $2$ respectively. We see that removing unrelaxed halos from the numerical catalogs of the different models recovers spin distributions which are well approximated by a log-normal and are consistent with each other over the same range of values.

\begin{figure}
\begin{tabular}{c}
\includegraphics[scale=0.4]{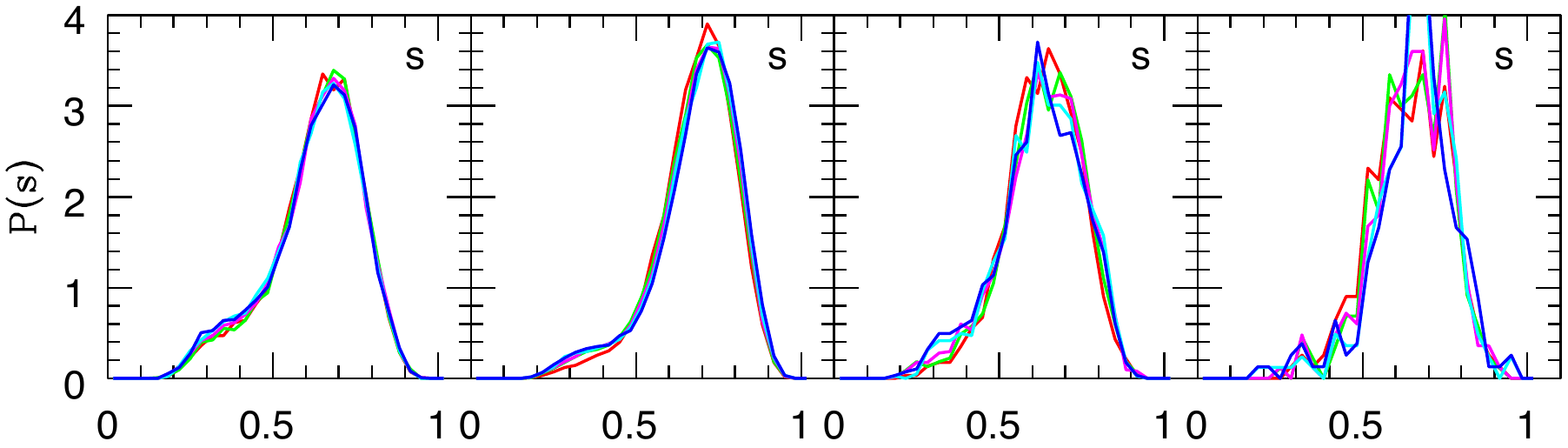}\\
\includegraphics[scale=0.4]{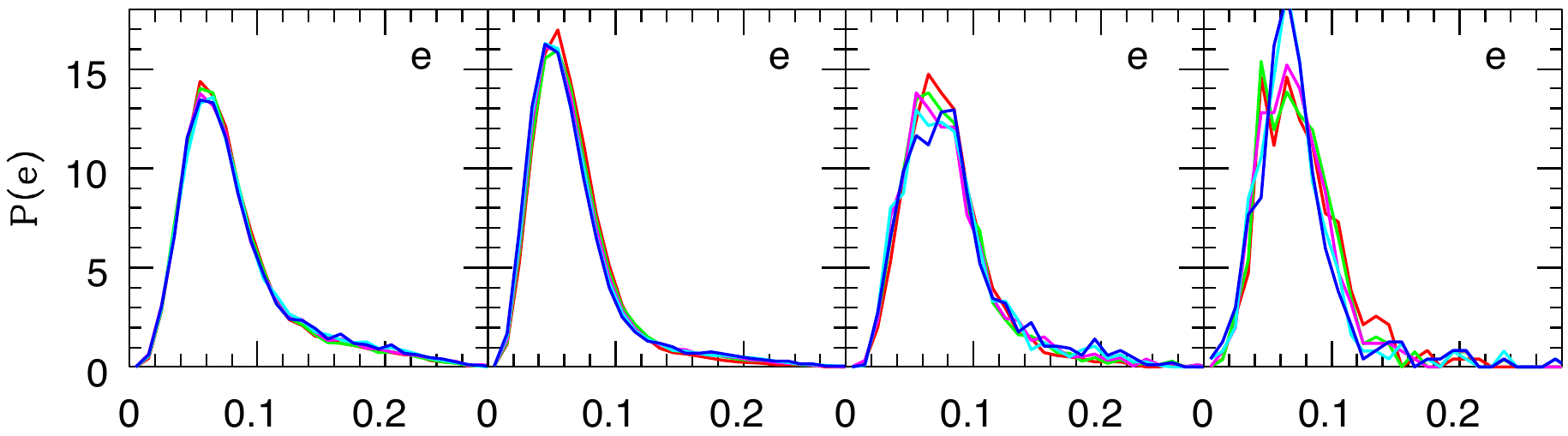}\\
\includegraphics[scale=0.4]{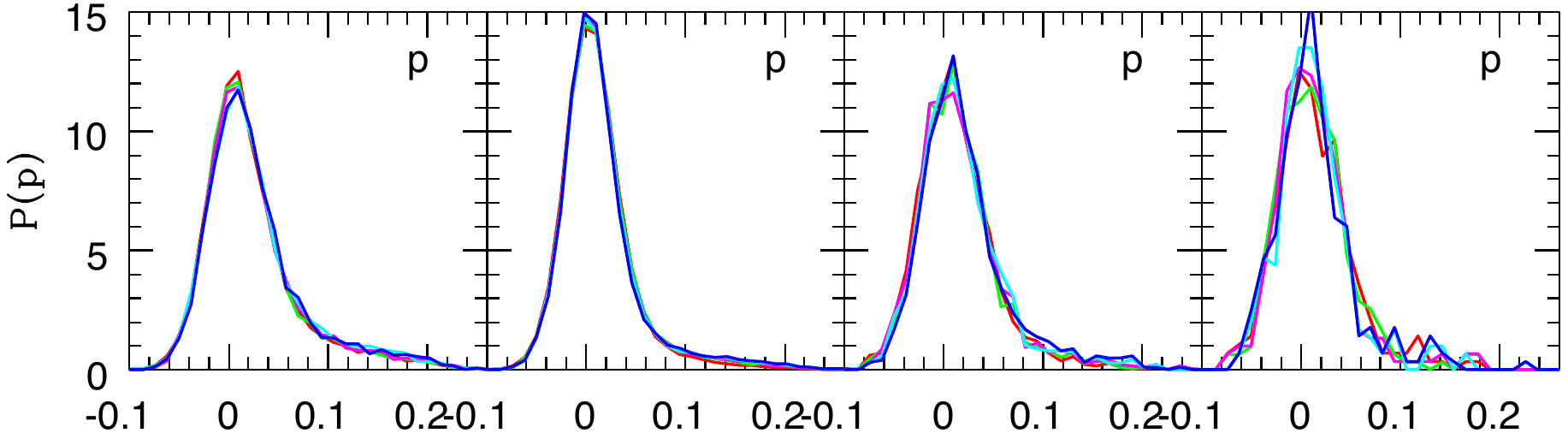}
\end{tabular}
\caption{Probability density function of halo sphericity, ellipticity and prolateness (panels top to bottom) at $z=0$ for $\Lambda$CDM, LFDM, WDM-a and WDM-b (panels left to right) respectively for halos with virial state parameter $0\leq\eta\leq1.5$. We consider the same mass bins as in Fig.~\ref{fig5}.}\label{fig9}
\end{figure}

In Fig.~\ref{fig9} we plot the normalized probability density function of the halo shape parameters after the cut $\eta>1.5$ for the same mass bins as shown in Fig.~\ref{fig5}. Again we see that having removed unrelaxed halos gives consistent distributions of the shape parameters at low masses with minimal contribution from objects with extreme values of ellipticity and prolateness. 


\section{Conclusions}\label{conclu}
Unphysical groups of particles arise from the artificial fragmentation of the matter density field in N-body simulations of non-standard DM models characterized by a small scale cut-off in the linear matter power spectrum. These spurious halos contaminate numerical halo catalogs altering the low-mass end of the halo mass function in the range of masses which are most relevant for observational tests of DM scenarios. In the case of Warm Dark Matter models these artifacts manifest as an upturn of the mass function at low masses. However, the absence of an upturn does not imply that the DM models with a softer suppression of power at small scales are exempt from spurious halo contamination, rather they are subdominant. 

Here, we have argued that the study of the structural properties of halos is a better proxy to differentiate between artificial group of particles and genuine halos. Using a suite of high-resolution simulations we have shown that spurious halos have systematically larger values of the spin parameter, are less spherical, more elliptical or prolate and strongly deviate from virial equilibrium compared to genuine halos. All these features are highly correlated for spurious halos and we find them even in the case of the LFDM scenario whose mass function does not exhibit any upturn at low masses. 

The strong correlation between the spin parameter and the deviation from virial equilibrium suggests that selecting approximately well relaxed halos can remove the bulk of spurious halo contamination. We find that a mildly conservative cut in the virial state parameter allows to recover corrected halo mass function. Most importantly, we show that this approach gives convergent results when applied to halo catalogs computed from N-body simulations with different mass resolution.

The advantage of this approach is twofold. One, it allows to estimate the correct halo abundance over a wider range of masses than a simple mass cut prescription which depends on the resolution of the simulation and the characteristics of the simulated cosmology. Two, as the halo selection is based on a physical criterion it is also applicable to cosmological models where the halo mass function lacks a visible upturn due to the spurious halos being subdominant at low masses.

\begin{acknowledgments}
The authors are grateful to Yann Rasera for useful discussion, and thank the referee for providing
constructive comments and help in improving the contents of this paper. This work was granted access to HPC resources of IDRIS through time allocations made by GENCI (Grand Equipement National de Calcul Intensif) on the machine ADA n.2014042287. The research leading to these results has received funding from the European Research Council under the European Community's Seventh Framework Programme (FP7/2007-2013 Grant Agreement no. 279954).
\end{acknowledgments}


\begin{thebibliography}{99}  
\bibitem{Zentner2003} A. R. Zentner, J. S. Bullock, Astrophys. J. {\bf 598}, 49 (2003)
\bibitem{Boyanovsky2008} D. Boyanovsky, H. J. de Vega, N. G. Sanchez, Phys. Rev. D {\bf 78}, 063546 (2008)
\bibitem{Hlozek2014} R. Hlozek, D. Grin, D. J. E. Marsh, P. Ferreira, arXiv:1410.2896
\bibitem{Das2011} S. Das, N. Weiner, Phys. Rev. D {\bf 84}, 123511 (2011)\bibitem{Maccio2012} A. V. Maccio {\sl et al.}, Mont. Not. Roy. Astron. Soc. {\bf 424}, 1105 (2012)
\bibitem{Vogelsberger2012} M. Vogelsberger, J. Zavala, A. Loeb, Mont. Not. Roy. Astron. Soc. {\bf 423}, 3740 (2012)
\bibitem{Lovell2012} M. R. Lovell {\sl et al.}, Mont. Not. Roy. Astron. Soc. {\bf 420}, 2318 (2012)
\bibitem{Zavala2013} J. Zavala, M. Vogelsberger, M. G. Walker, Mont. Not. Roy. Astron. Soc. {\bf 431}, L20 (2013)
\bibitem{Anderhalden2013} D. Anderhalden {\sl et al.}, J. Cosmol. Astropart. Phys. {\bf 03}, 014 (2013)
\bibitem{Lovell2014} M. R. Lovell {\sl et al.}, Mont. Not. Roy. Astron. Soc. {\bf 439}, 300 (2014)
\bibitem{Abazajian2014} K. N. Abazajian, Phys. Rev. Lett. {\bf 112}, 161303 (2014)
\bibitem{Schneider2014a} A. Schneider, D. Anderhalden, A. V. Macci\`o, J. Diemand, Mont. Not. Roy. Astron. Soc. {\bf 441}, L6 (2014)
\bibitem{Boehm2014} C. B\oe{}hm {\sl et al.},  Mont. Not. Roy. Astron. Soc. {\bf 445}, L31 (2014)
\bibitem{Buckley2014} M. R. Buckley {\sl et al.}, Phys. Rev. D {\bf 90}, 043524 (2014)
\bibitem{Agarwal2014} S. Agarwal, P.S. Corasaniti, S. Das, Y. Rasera, arXiv:1412.1103 
\bibitem{Gotz2002} M. G\"{o}tz, J, Sommer-Larsen, Ap\&SS {\bf 281}, 415 (2002)
\bibitem{Gotz2003} M. G\"{o}tz, J, Sommer-Larsen, Ap\&SS {\bf 284}, 341 (2003)
\bibitem{WangWhite2007} J. Wang, S. D. M. White, Mont. Not. Roy. Astron. Soc. {\bf 380}, 93 (2007) 
\bibitem{BoylanKolchin2011} M. Boylan-Kolchin, J. S. Bullock, M. Kaplinghat, Mont. Not. Roy. Astron. Soc. {\bf 415}, L40 (2011) 
\bibitem{BoylanKolchin2012} M. Boylan-Kolchin, J. S. Bullock, M. Kaplinghat, Mont. Not. Roy. Astron. Soc. {\bf 422}, 1203 (2012)
\bibitem{Zavalaetal2009} J. Zavala {\sl et al.}, Astrophys. J. {\bf 700}, 1779 (2009)
\bibitem{Papastergis2012} E. Papastergis, A. M. Martin, R. Giovanelli, M. P. Haynes, Astrophys. J. {\bf 739}, 38 (2012)
\bibitem{Klypin2014} A. Klypin, I. Karachentsev, D. Makarov, O. Nasonova, arXiv:1405.4523
\bibitem{Schneider2013} A. Schneider, R. E. Smith, D. Reed, Mont. Not. Roy. Astron. Soc. {\bf 433}, 1573 (2013) 
\bibitem{Hahn2013} O. Hahn, T. Abel, R. Kaehler, Mont. Not. Roy. Astron. Soc. {\bf 434}, 1171 (2013) 
\bibitem{Angulo2013} R. E. Angulo, O. Hahn, T. Abel, Mont. Not. Roy. Astron. Soc. {\bf 434}, 3337 (2013)
\bibitem{Schneider2014b} A. Schneider, arXiv:1412.2133
\bibitem{Teyssier2002} R. Teyssier, Astron. \& Astrophys. {\bf 385}, 337 (2002)
\bibitem{Guillet2011} T. Guillet, R. Teyssier, Journ. Comp. Phys. {\bf 230}, 4756 (2011)
\bibitem{Viel2013} M. Viel, G. D. Becker, J. S. Bolton, M. Haehnelt, Phys. Rev. D {\bf 88}, 043502 (2013) 
\bibitem{Lewis2000} A. Lewis, A. Challinor, A. Lasenby, Astrophys. J. {\bf 538} 473 (2000)
\bibitem{Bode2001} P. Bode, J. P. Ostriker, N. Turok, Astrophys. J. {\bf 556} 93 (2001)
\bibitem{Prunet2008} S. Prunet {\sl et al.}, Astrophys. J. Supp. {\bf 178}, 179 (2008)
\bibitem{Crocce2006} M. Crocce, S. Pueblas, R. Scoccimarro, Mont. Not. Roy. Astron. Soc. {\bf 373}, 369 (2006)
\bibitem{Roy2014} F. Roy, V. Bouillot, Y. Rasera, Astron. \& Astrophys. {\bf 564}, A13 (2014)
\bibitem{Davis1985} M. Davis, G. Efstathiou, C. S. Frenk, S. D. M. White, Astrophys. J. {\bf 292}, 371 (1985)
\bibitem{Schneider2012} A. Schneider, R. E. Smith, A. Macci\'o, B. Moore, Mont. Not. Roy. Astron. Soc. {\bf 424}, 684 (2012)
\bibitem{Bullock2001} J. S. Bullock {\sl et al.}, Astrophys. J. {\bf 555}, 240 (2001) 
\bibitem{Thomas1998} P. A. Thomas {\sl et al.}, Mont. Not. Roy. Astron. Soc. {\bf 296}, 1061 (1998)
\bibitem{Bett2007} P. Bett {\sl et al.}, Mont. Not. Roy. Astron. Soc. {\bf 376}, 215 (2007)
\bibitem{Knebe2008} A. Knebe, C. Power, Astrophys. J. \textbf{678}, 621 (2008)
\bibitem{Balmes2014}  I. Balmes, Y. Rasera, P.S. Corasaniti and J.-M. Alimi, Mont. Not. Roy. Astron. Soc. {\bf 437}, 2328 (2014), arXiv:1307.2922
 

\end{thebibliography}
\end{document}